\documentstyle{aipproc}

\input psfig                       

\def \gray     {$\gamma$-ray }
\def \grays    {$\gamma$-rays}

\def \sig      {$\sigma$ }
\def \sg1      {$\sigma$}
\def \~        {$\sim$}

\begin{document}

   \title{Evidence for $\gamma$-Ray Flares in 3C 279 and PKS 1622-297
          at $\sim$10~MeV}
   \author{W. Collmar$^1$, V. Sch\"onfelder$^1$,
           H. Bloemen$^2$, J.J. Blom$^2$, W. Hermsen$^2$,
           M. McConnell$^3$, J.G. Stacy$^3$,
           K. Bennett$^4$, O.R. Williams$^4$}
   \address{
         $^1$Max-Planck-Institut f\"ur Extraterrestrische Physik,
             85740 Garching, Germany \\
         $^2$SRON-Utrecht, Sorbonnelaan 2, 3584 CA Utrecht, The Netherlands \\
         $^3$University of New Hampshire, IEOS, Durham NH 03824, USA \\
         $^4$Astrophysics Division, SSD/ESA, NL-2200 AG Noordwijk,
            The Netherlands }
   \maketitle

\begin{abstract}
The EGRET experiment aboard the Compton Gamma-Ray Observatory (CGRO) has 
observed at energies above 100 MeV strong gamma-ray flares with short-term
time variability from the gamma-ray blazars 3C~279 \cite{Wehrle97} and
PKS~1622-297 \cite{Mattox97}. During these flaring periods both
blazars have been detected by the COMPTEL experiment 
aboard CGRO at photon energies of $\sim$10~MeV, revealing
simultaneous \gray activity down to these energies.
For both cases the derived fluxes exceed those measured in 
previous observations, and 3C~279 shows an indication for 
time variability within the observational period.
Both sources show evidence for `hard' MeV spectra.
In general the behaviour of both sources at \gray energies is found
to be quite similar 
supporting the conclusion that the underlying physical mechanism for 
both \gray flares might be the same.  
\end{abstract}

\section*{Introduction}
The EGRET experiment aboard CGRO has detected more than 60 blazar-type
AGN \cite{Kanbach96} thereby greatly widening the field of extragalactic \gray
astronomy. Most of them are observed to be time variable and several sources
showed remarkable flares. During the last two years the two most intense 
flares along the whole EGRET mission have been observed from the 
sources 3C~279 \cite{Wehrle97} and PKS~1622-297 \cite{Mattox97}, 
which occured on top of an already high \gray flux level. 

The COMPTEL experiment \cite{Schoenfelder93}, measuring 
0.75-30~MeV \grays, has detected 8 of these \~ 60 EGRET blazars 
\cite{Collmar96}. Among them are 3C~279 and PKS~1622-297.
In this paper we report first results on these sources 
for the time periods for which these flares have been observed
by EGRET at energies above 100~MeV.      

\section*{Observations and Data Analysis}
The \gray flare events have been observed during a three week observation 
of the Virgo sky region from January 16 to February 6, 1996 for 3C~279 in CGRO Cycle~5 during the viewing periods (VPs) 511.0 and 511.5, and for PKS~1622-297
during a four week observation towards the Galactic Center region
from June 6 to July 10, 1995 in CGRO Cycle~4 covering the VPs 421-423.5. 

We have applied the standard COMPTEL maximum-likelihood analysis method
(e.g. \cite{Boer92}) to derive detection significances, fluxes, and 
flux errors of \gray sources in the four standard COMPTEL energy bands (0.75-1~MeV, 1-3~MeV, 3-10~MeV, 10-30~MeV), and a background modelling
technique which eliminates in a first approximation source signatures but preserves the 
general background structure \cite{Bloemen94}. 
For the 10-30~MeV range the improved COMPTEL data cuts \cite{Collmar97},
increasing its' sensitivity, have been applied. 
To derive source fluxes, several sources located in the surrounding sky region 
(e.g. 3C~279 and 3C~273 in Virgo) have been simultaneously fitted in an
iterative procedure, leading to a simultaneous determination of the fluxes of several potential sources and a background model which takes into
account the presence
of sources. For the analysis of PKS~1622-297 a diffuse emission model
has been included in the fitting procedure as well.

\section*{Results}
\subsubsection*{3C~279}
The blazar 3C~279 is detected by COMPTEL with a significance
of \~ 4\sig during 
this observational period of three weeks on Virgo in CGRO Cycle~5 (Fig. 1).
The observed flux level in the 10-30~MeV band is the highest ever detected. 
This is the first redetection of 3C~279 by COMPTEL for energies above 
($>$3~MeV) since 1991, when the blazar showed another \gray flare 
observed simultaneously by EGRET and COMPTEL 
($\!$\cite{Kniffen93,Williams95}). 
Subdividing the three week period into the individual
VPs 511.0 (two weeks) and 511.5 (one week) reveals evidence for a flux jump by roughly a factor of 4 (Fig. 2) within 10~days. 
The flux value of 3C~279 measured in VP 511.5 is the largest ever
observed by COMPTEL. The significance that the two fluxes are different
is 2.6$\sigma$, and represents the shortest time varibility yet observed 
by COMPTEL from any blazar.     
This rise in  
flux is consistent with the EGRET observations at energies above 100~MeV. 
During VP~511.0 3C~279 was redetected by EGRET at a high flux level 
and rose up to the largest value ever in VP~511.5 \cite{Wehrle97}.

\def\bbllx{ 2.0cm}
\def\bblly{ 8.5cm}
\def\bburx{19.5cm}
\def\bbury{25.9cm}

\begin{figure} [t]
\centering{
\psfig{figure=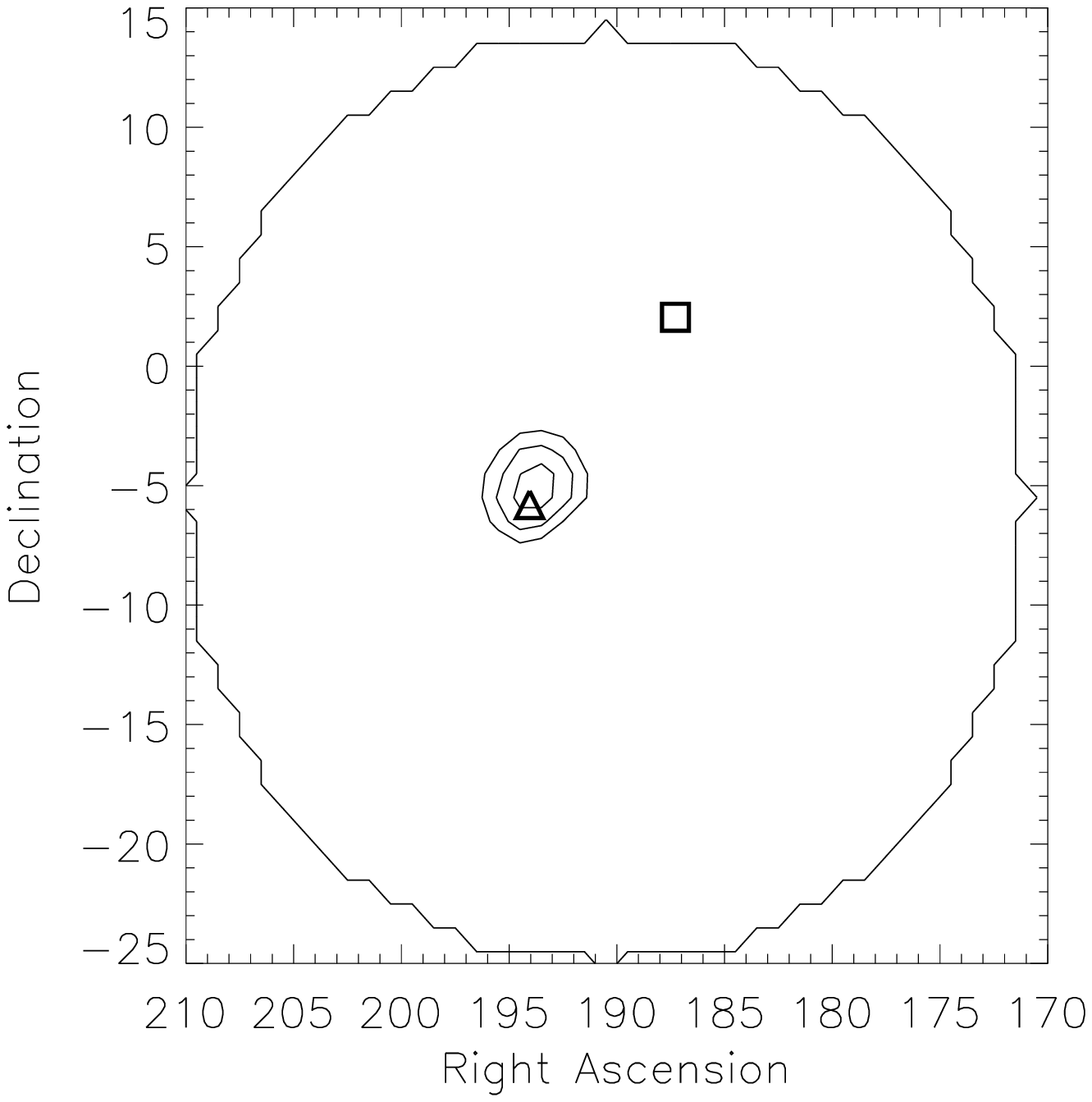,width=7.1cm,height=7.5cm}
\vspace*{-7.53cm}\hfill
\psfig{figure=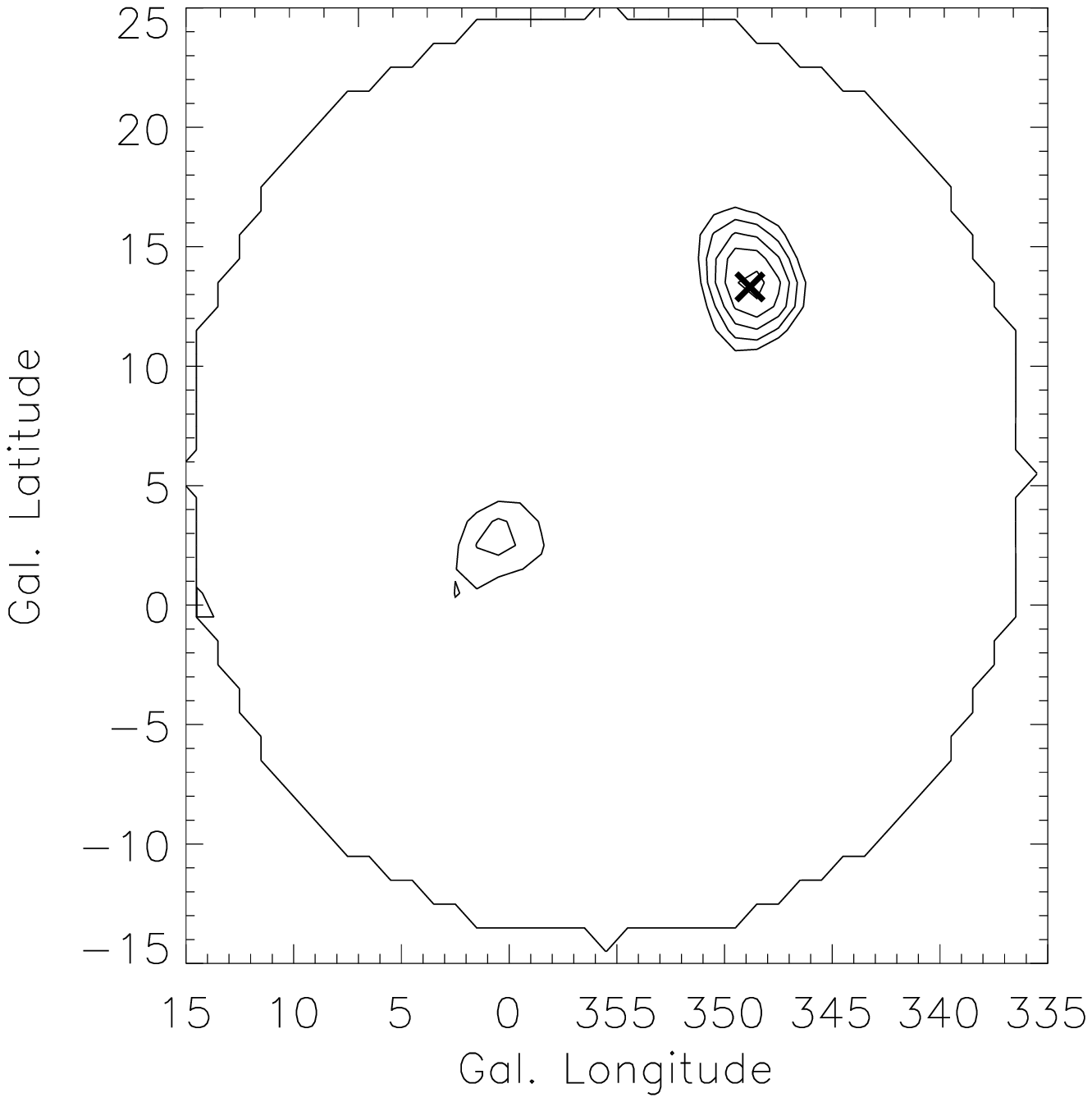,width=7.1cm,height=7.5cm}
   }
\vspace{0.3cm}
\caption[]{
COMPTEL 10-30~MeV skymaps (detection significances) 
of the Virgo region (left) and the galactic center (right) for the relevant
time periods. The contour lines start at a detection significance of 3.0\sig with a step of 0.5\sig assuming $\chi^2_1$-statistics for known sources.
The locations of 3C~279 ({\bf $\Delta$}), 3C~273 ({\bf $\Box$}), and PKS~1622-297 ({\bf X}) are indicated.
}
\end{figure}
%

The spectral analysis shows positive evidence for the source only 
at energies above 3~MeV. Together with the upper limits derived at the lower
energy bands, this indicates a `hard' (photon spectral
index $\alpha<$2) energy spectrum at MeV energies (Fig.~3).      

\begin{figure}[t]
 \psfig{figure=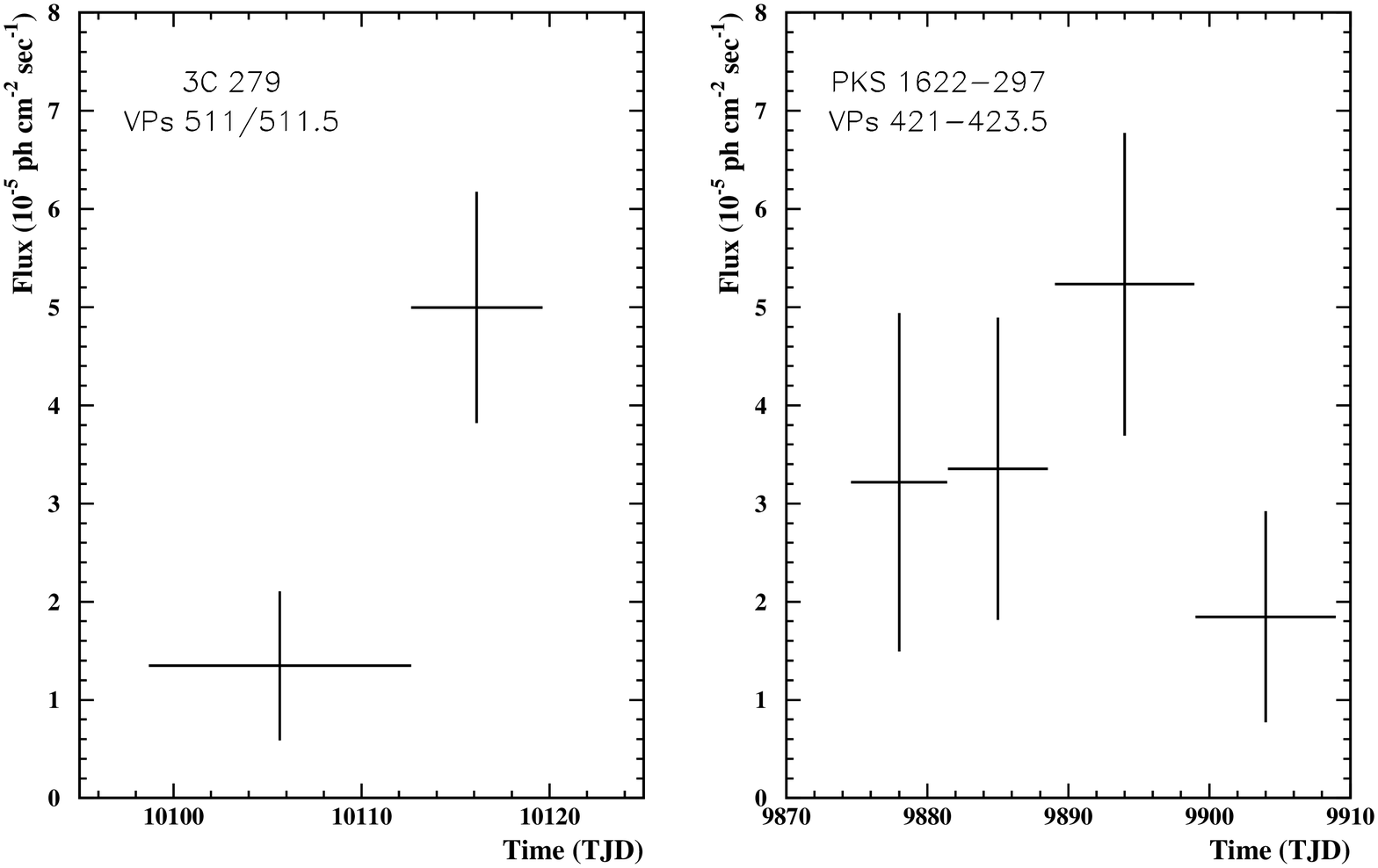,height=8cm,width=13.8cm}
 \vspace{-0.2cm}
 \caption[]{
Light curves in the 10-30~MeV range of 3C~279 (left) and PKS~1622-297
(right). The fluxes are given for the individual CGRO VPs. The error bars 
are 1\sg1 . The flux change of 3C~279 by a factor of 3.7 within 10~days 
is the shortest time variability yet observed from any blazar by COMPTEL. 
The fluxes measured during these flaring intervals are the largest ever
observed from any blazar by COMPTEL in this energy band.   } 
\end{figure}

\subsubsection*{PKS~1622-297}
The blazar PKS~1622-297 is detected with a significance of \~ 5\sig  during
the four week pointing towards the Galactic Center in CGRO Cycle~4 (Fig.~1).
This is the first detection of this blazar by COMPTEL as was the case
for EGRET \cite{Mattox97}.
The COMPTEL light curve, even though the flux variations are not statistically significant, follows the general trend reported by EGRET at energies 
above 100~MeV. There is evidence for the source during all four individual VPs
showing MeV flaring of PKS~1622-297 for at least one month (Fig.~2). 
The largest flux value is observed during VP~423.0 consistent with the time
period of the major flare observed by EGRET.  
A flux drop by a factor of \~ 2.5 between the two
last VPs, again consistent in trend with EGRET, is a hint
for MeV variability as well. 

The spectral analysis shows that the source is mainly detected in the 
highest COMPTEL energy band (10-30~MeV), which, together with the 
upper limits derived at lower energies, indicates a
'hard' (photon spectral index $\alpha<$2) spectrum on average (Fig.~3).
However, we like to point out that 
the source is located just above the plane near the galactic center region, 
which is a difficult region for quantitative analysis, especially for energies 
below 3~MeV, due to the diffuse MeV emission of the Galaxy.
Although a diffuse emission model was included in the analysis procedure,
the presented spectral results should be considered as preliminary.

\begin{figure}[t]
 \psfig{figure=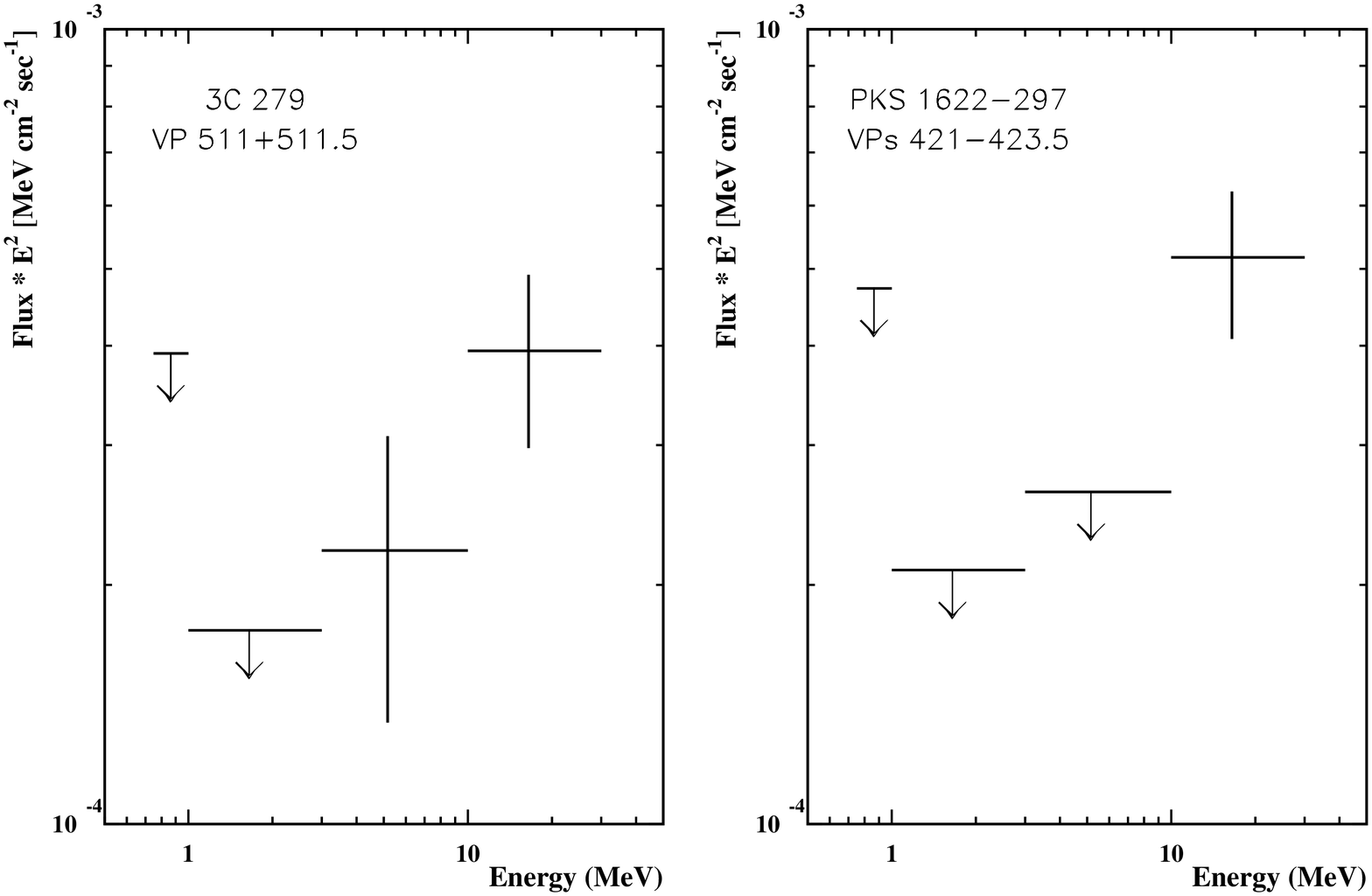,height=8cm,width=13.8cm}
 \vspace{-0.2cm}
 \caption[]{
Energy spectra of 3C~279 (left) and PKS~1622-297
(right) are shown. The spectra, averaged over the whole 3- and 4 week
observations of both blazars, are presented in a differential
flux $\times$ E$^2$ representation. 
The error bars are 1\sig and the upper limits are 2\sg1 . 
Both sources are mainly detected at the upper COMPTEL energy bands, which, 
together with the upper limits at lower energies, indicates a 'hard' 
MeV spectrum for both cases. 
  }
\end{figure}

\section*{Summary}
We have reported first results of COMPTEL observations of two blazars,
3C~279 and PKS~1622-297, for time periods for which EGRET ($>$100~MeV)
observed the two strongest \gray flares ever occuring on top of an already 
high flux level.
Although there are differences in detail, the general MeV behaviour of both 
sources resembles each other surprisingly accurately. Both sources are 
detected during these periods of high \gray activity at 
high \gray energies, which by itself demonstrates simultaneous  
MeV-flaring activity. Note, that PKS~1622-297 is detected for the first
time by COMPTEL. Both sources are only detected at the highest
COMPTEL energies. This, together with the upper limits derived at the lower 
energies, leads to evidence for hard (photon spectral index
$\alpha<$2) MeV spectra.
The 10-30~MeV flux follows for both sources the general flux trend 
as seen by EGRET with evidence for time variability in the case 
of 3C~279 and a hint in the case of PKS~1622-297.    
These similarities support the conclusion that the underlying physical 
mechanism for \gray activity is the same for both sources. 

Detailed COMPTEL analyses, concentrating on subsets of these observations
are in progress to derive 
informations on possible time-shifts between the high-energy EGRET and the 
low-energy COMPTEL \gray emission. Especially PKS~1622-297 is a promising candidate for such investigations because the observation covers a time 
period of high \gray activity in which a flux spike is observed.
For 3C~279 the observations are only available at the rising part of the
flare. It remains unclear whether the top of the MeV emission is covered
by COMPTEL.

\end{document}